# Secure and Image Retrieval based on Multipurpose Watermarking for Mammography Images Database

H. Ouahi
Laboratory LabSiv
University Ibn Zohr of Agadir FSA BP 28/S
Morocco

M. El Hajji
IRF-SIC Laboratory
University Ibn Zohr of Agadir FSA BP 28/S
Morocco

K. Afdel
Laboratory LabSiv
University Ibn Zohr of Agadir FSA BP 28/S
Morocco

## ABSTRACT
In the cancerology domain, we were brought to make periodic mammography images to monitor tumor patients. Oracle Database Management system (DBMS) is a solution to manage these images with patient's data recorder. Knowing the large size of medical images of mammograms, the Oracle DBMS saves these images outside the Oracle database using external LOBs. The link between these images and Oracle is done through the BFILE. At this level, two problems are raised: the first problem is that access to these images can become impossible because the link is likely to be broken. The second problem is security, the fact that the images are saved outside the Oracle database, they do not benefit from its powerful security. The protection of the integrity and confidentiality of data and patient images are a necessity defended by laws and they must be preserved against any unauthorized access, alteration or destruction. In this paper, we propose the method of reversible watermarking technique based on the difference expansion to resolve these two problems and explore its use in search and retrieval strategy of images.

## 1. INTRODUCTION
As for as the cancerology domain is concerned, doctors are likely to make periodic mammography images to monitor tumor patients. For the management of these images several databases image have been developed in terms of the size and variety. In recent years, various search engines based image search by content (CBIR) [10][6][13][12][15] have been developed by different research groups. Generally, these systems use CBIR retrieval algorithms based on vectors descriptions of low-level features such as texture, color and shape. The computation time of these algorithms is important to extract an image that is similar to the query image. The reason is that we need to calculate the vectors of features descriptions of all the images in the database and compare them to the vector description characteristics of the query image so as to keep only the closest one. Management system oracle database (DBMS) provides the solution ORD Image signature by doing the calculating and recording the characteristics of the describing vector (signatures) for each image in the database. On the one hand, the parameters of the describing vector are not well suited for medical images. On the other hand, given the large size of medical images of mammograms, the Oracle DBMS record these images outside of the Oracle database. The link between these images and LOBs in Oracle is done through a pointer BFILE. At this level, there are two problems: the first one is concerned with the breakable relationship between LOBs and images. The second problem is that the images that are stored outside the Oracle database do not benefit from its powerful security. Therefore one of the security measures that can be used is the watermarking [8][11][3]. The latter is an important area of research for the security, confidentiality and integrity of data. Watermarking can be done by hiding the electronic patient record in its medical images. Thus, the image and patient data become a single entity. However, in medical diagnostic, quality is very critical and must not be compromised. Therefore, any change in the content of the image is not tolerated or permitted. For this reason, we adopted the method of fragile and reversible watermarking technique using difference expansion [7] [4].

In this paper, we define in Section 2 suitable attributes characterizing for medical images. These attributes will be embedded in all images of the medical database by using watermarking. In Section 3, we present reversible watermarking method based on the difference expansion. In Section 4, we will explore the use of watermarking in the search and retrieval images in a watermarked database images. In addition to this, we are going to restore the broken external LOBs by using watermarking. In section 5, experimental results are given. The conclusion is described in the last section.

## 2. EXTRACTION CHARACTERISTICS OF IMAGE
In medical images, there is the problem of choosing a set of relevant characteristics. In the context of medical imaging, a study was carried out to define the relevant attributes characterizing the medical images [2]. These attributes are

- Two textures attributes: "difference variance" and "difference entropy" derived from the co-occurrence Haralick matrix[1];

- The statistical moments of the image gray levels: The "Average", "Standard Deviation", "Skewness" and "Kurtosis";

- The mean values of the moments of order 1, 2, 3 and 4, which are invariant to the image. These last moments are widely used in pattern recognition to describe information related to the shape of objects in the image.

## 3. THE METHOD OF REVERSIBLE WATERMARKING BY DIFFERENCE EXPANSION

Medical images, unlike most images require special attention. The integration of more data in its content should not affect the quality and readability of the image in order to avoid bad diagnosis which can be caused by this integration. We propose reversible watermarking method based on the difference expansion. This method allows finding exactly the original image (without distortion), after the extraction of the embedded information.

### 3.1 Description of the proposed method

In the Watermarking method based on the difference expansion, the mark or payload is embedded into LSB bits of the difference value. If we take a pair of grayscale value of pixel (x, y); x,y ∈ Z, (0 <= x,y <= 255),

The average value is given by: $l = \lfloor \frac{x+y}{2} \rfloor$

and the difference is given by : $h = x-y$

The symbol $\lfloor . \rfloor$ represents the integer value less than or equal. To recover the original value of the pixel, we simply need to reverse the above equations:

$x = l + \lfloor \frac{h+1}{2} \rfloor$ , $y = l - \lfloor \frac{h}{2} \rfloor$

Knowing the values of gray levels of the pixels must belong to the interval [0,255], it is necessary that

$(0 <= l + \lfloor \frac{h+1}{2} \rfloor <= 255, 0 <= l - \lfloor \frac{h}{2} \rfloor <= 255)$

These two inequalities are equivalent to

$|h| <= \min(2(255-l), 2l+1)$

During the mark integration in the image, the pixels of the image are grouped into three zones according to the difference value of h: the extensible zone, changing zone and the unchanging zone. The data bits of the payload are included only in extensible and changeable zones [5]. The Location map bits integration is established to distinguish these three zones.

### 3.2 Data embedding in image

The data embedding algorithm consists of five steps:

- Calculating the difference values and average;
- Partitioning difference values into five sets;
- Creating a location map;
- Collecting original LSB values;
- Data embedding by replacement or adding LSB bits;

Figure 1 shows the data embedding algorithm in the image:

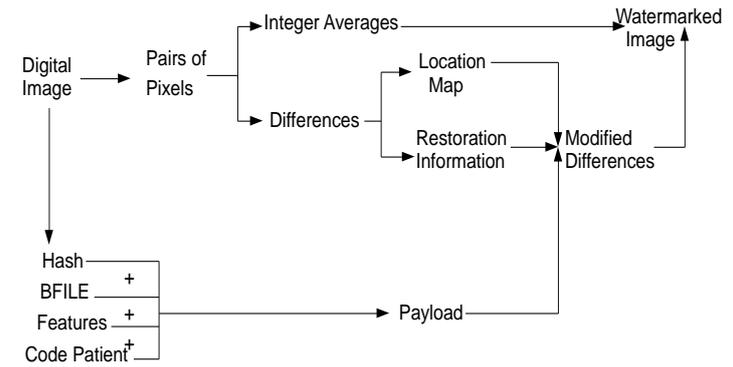

**Fig 1: Payload embedding Algorithm**

Our system offers multipurpose watermarking composed with four parameters which are:

- **Hash**: used to authenticate the watermarked image and protect the private information of the patient;
- **BFILE**: used to restore the broken relationship between patient data in the Oracle database and its images located outside of the database;
- **Features Vector**: A vector consisting of seven values (difference variance, difference entropy, mean, standard deviation, skewness, kurtosis,

  the mean moments of order 1, 2,3 and 4) these parameters will be used to improve extracting images from the watermarked database;
- **Patient code**: will be used to link the electronic patient record and its medical images.

### 3.3 Extracting data from the watermarked image

The decoding and authentication algorithm consists of five steps:(figure 2)

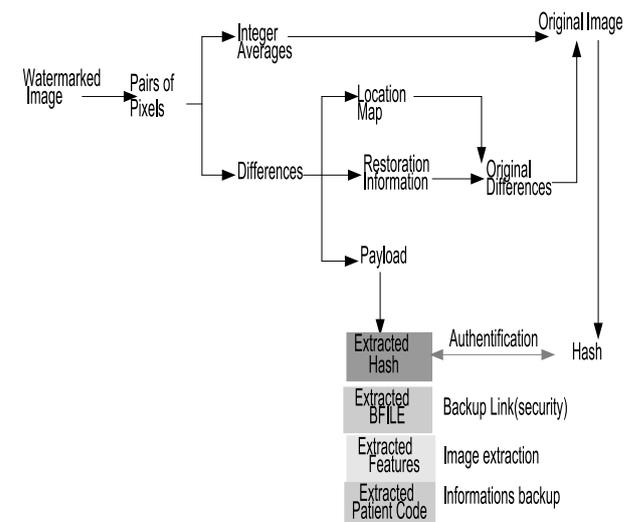

**Fig 2: Payload extraction Algorithm**

- Calculating difference values and average;
- Creating two disjoint sets of difference values: changeable and not changeable;
- Collecting LSBs of all difference values in changeable;
- Decoding the location map from LSBs and restoring the original values of differences;
- Content authentication and original content restoration.

### 3.4 The offline watermarking database image

In this work, we use the mammography image database "MIAS"[14]. Indeed, this medical image database is considered as benchmark in this field. The database contains the images of the left and the right breast of 161 patients. The database contains 322 images.

To improve the search and retrieval images, our system proposes using watermarking techniques. In order to do this, and before performing online search and retrieval images, we should first embed payload in each image of the database. This payload is composed from four parameters (Hash, BFILE, Feature Vector and Code Patient) figure 3.

## 4. IMAGE SEARCH AND RETRIEVAL BASED WATERMARKING

After embedding payload in all images of the database, we can perform the online search and retrieve images. In order to protect patient's data, the system starts by checking the authentication hash parameter before starting the search and retrieval. Our system provides two kinds of queries:

1. The first one is based on the patient's code to extract images and its electronic data recorder, which allows doctors to monitor the tumor over time and avoid serious problems, caused by the loss and permutation of the mammography images.

2. The second is based on the image characteristic features. The system will first calculate online the features of the query image. Then, these features are compared with the features extracted from watermarked image database in order to retrieve similar images. Here, Euclidean distance is employed as the similarity measure

The parameter of payload "BFILE" is used only in the case of broken links made by external LOBs between the data and images. The system can restore these links based on the BFILE extracted from each image of the watermarked database images.

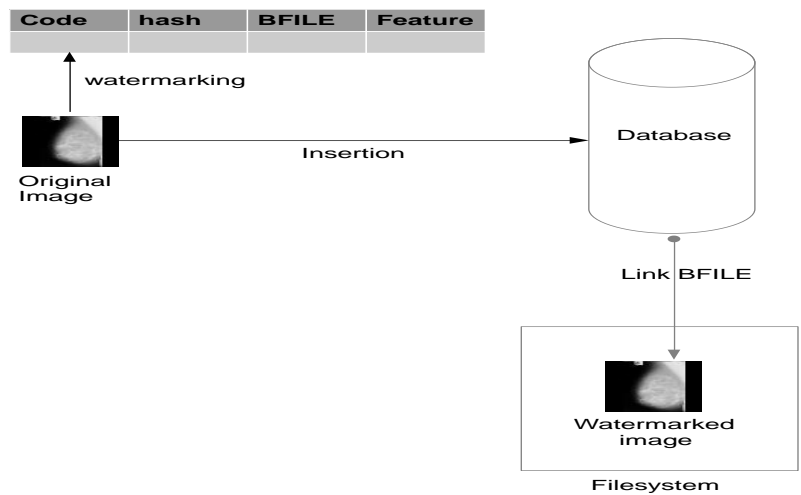

**Fig 3: offline watermarking**

## 5. EXPERIMENTS AND RESULTS
### 5.1 The watermarking results
The Peak Signal to Noise Ratio (PSNR) is used as distortion measurement between the original and a watermarked image. It is define as

$$PSNR = 10\log_{10}\left(\frac{255^2}{1/N \sum_{n=1}^{N} w_n^2}\right)$$

Where N is the total number of the pixels, $w_n = s_n - y_n$, s represents the original image, and y is the watermarked image.

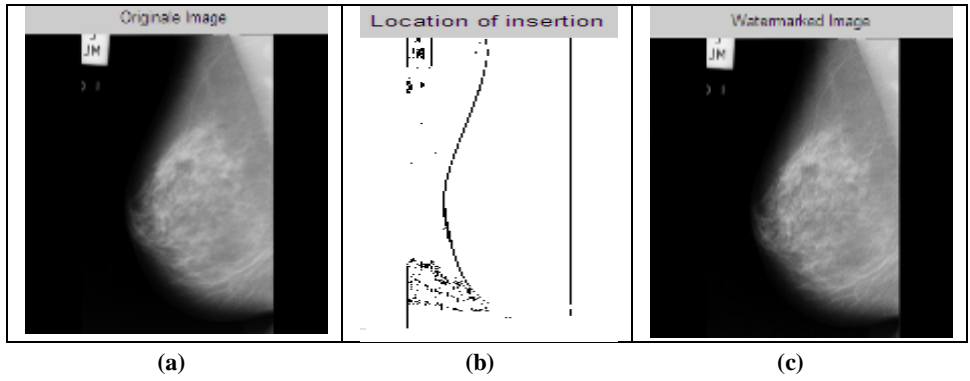

**(a)** **(b)** **(c)**
**Fig 4: (a) Original image (b) location of insertion (c) watermarked image(PSNR=44,19)**

And

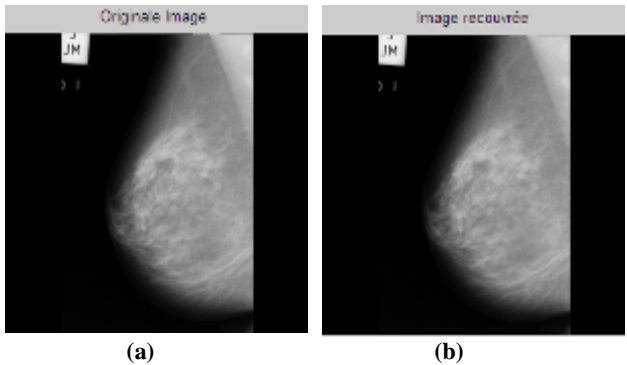

**(a)** **(b)**
**Fig 5: (a) Original image (b) recovered image (PSNR=71,69)**

At figure4 and figure5 we can see the difference between the original, the watermarked and the recovered image.

## 5.2 Image retrieval performance
We perform the experiments based on a test database with 63 images in 3 classes, each class includes 21 images. The size of these images is: 256× 256.
For the query based on features, we show an example of retrieval results in Figure 6.

From the figure 6, we can see that during the online retrieval process, content integrity can be verified simultaneously (hash of image) if the image is suspected to be attacked or tampered. In order to backup the breakable link between database and images, one should rebuild the BFILE following the steps previously described and update the database.  a simple click on the button regeneration BFILE permits to do that.
We use the Euclidean distance (D) between query image and the database images.
For the feature-based query scheme, we compute the feature online for the query image, then we compare it with the feature watermark in each database image in order to find the most desired similar images.
The recall and precision parameters for each class is shown in table1.

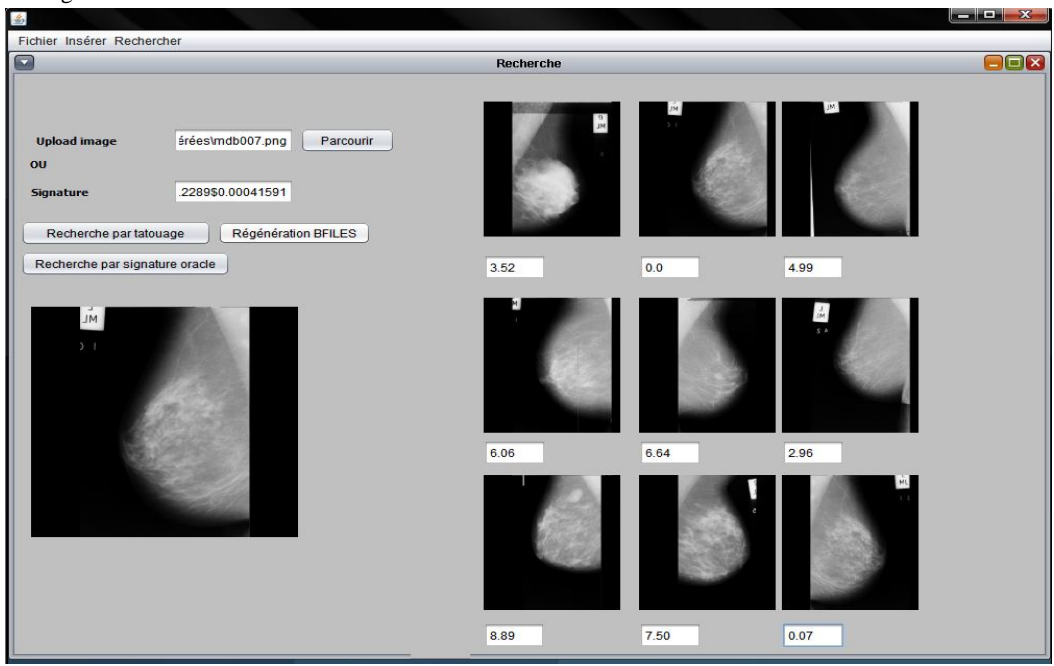

**Fig 6: Image retrieval**

**Table1: Mean recall/precision of class**

| D | Mammography | | Bone | | Cardiac | |
|---|---|---|---|---|---|---|
| | Recall | Precision | Recall | Precision | Recall | Precision |
| 1 | 0,05 | 1,00 | 0,05 | 1,00 | 0,05 | 1,00 |
| 5 | 0,09 | 0,77 | 0,05 | 0,73 | 0,07 | 1,00 |
| 10 | 0,31 | 0,77 | 0,10 | 0,76 | 0,14 | 0,69 |
| 20 | 0,62 | 0,68 | 0,31 | 0,46 | 0,35 | 0,69 |
| 40 | 0,97 | 0,60 | 0,72 | 0,40 | 0,71 | 0,61 |

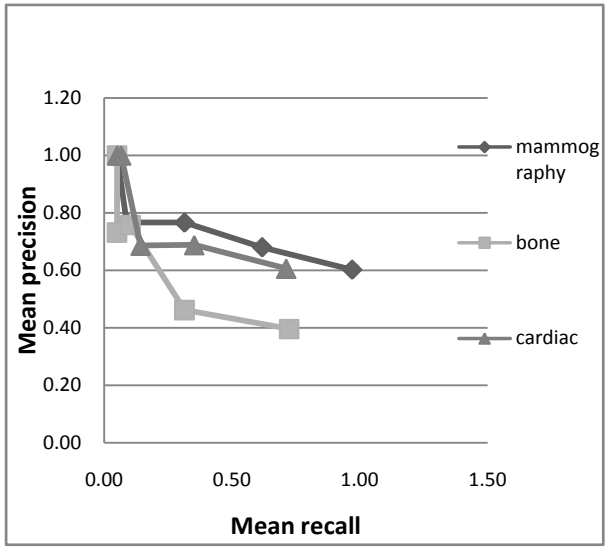

**Fig 7 : Precision/Recall graph**

The precision-recall (PR) graphs [9] shown in Fig. 7 is plotted using the mean precision and recall rate calculated from the requests made by the choice of five random images of each three classes and five distance.

From the graph above, we can see that for the distance 1 we have a maximum average precision 1, and minimum average recall 0.05 for the three classes. At the distance 20 the two values are good for the images of the mammography, which is the opposite for bone and cardiac images. At the distance 40, we discover that the average precision and average recall are specific to the mammography class.

As a conclusion to this result, our system with the watermarking technique used and the features selected will make the user satisfied, because it will not waste time reading uninteresting information (average precision) and have access to all information that he wanted (recall correctly).

## 6. CONCLUSION
In this paper, we use the multipurpose watermarking composed with four following parameters. The objective of this is to enhance the safety management system of the patient's data recorder and medical images using Oracle DBMS. The advantages of this system reside in three aspects. First, the system begins to check the image authentication. Second, it allows the search and retrieval of images based on their characteristics features embedded in images. Third, the fact that the images are saved outside of the Oracle database DBMS, the system restores broken links between these images and LOBs in Oracle based on the BFILE extracted from the watermarked image. The perspective of this work is to apply this technique for medical examination videos.

## 7. ACKNOWLEDGMENTS
The authors would like to thank the anonymous reviewers for their valuable comments and suggestions to improve the quality of the paper.